\documentclass[runningheads]{llncs}
\usepackage[T1]{fontenc}
\usepackage{graphicx,verbatim}
\usepackage{url}
\usepackage{xspace}
\usepackage{amsmath}
\usepackage{booktabs}
\usepackage[misc]{ifsym}  % corresponding author
\usepackage[colorlinks=true,linkcolor=blue,citecolor=blue,urlcolor=blue]{hyperref}
\usepackage{color}
\usepackage{xcolor}

\usepackage[capitalize,noabbrev]{cleveref}
\usepackage{appendix}
\usepackage{cleveref}

\crefname{figure}{Fig.}{Figs.}

\crefname{section}{Section}{Sections}
\crefname{theorem}{Theorem}{Theorems}
\crefname{lemma}{Lemma}{Lemmas}
\crefname{equation}{Equation}{Equations}
\crefname{proposition}{Proposition}{Propositions}
\crefname{claim}{Claim}{Claims}
\crefname{appendix}{Appendix}{Appendices}
\crefname{algorithm}{Algorithm}{Algorithms}
\crefname{figure}{Figure}{Figs}
\crefname{table}{Table}{Tables}
\crefname{remark}{Remark}{Remarks}
\crefname{definition}{Definition}{Definitions}
\crefname{equation}{Equation}{Equations}
\crefname{corollary}{Corollary}{Corollaries}
\crefname{section}{Method}{Methods}

\newcommand{\eg}{\textit{e.g.}\xspace}
\newcommand{\ie}{\textit{i.e.}\xspace}

\newlength\savewidth

\newcommand{\mytask}{ICG\xspace} 
\newcommand{\datasetname}{ICG-CXR\xspace}
\newcommand{\methodname}{ProgEmu\xspace}
\newcommand{\methodnamesep}{ProgEmu\textsuperscript{2}\xspace}

\begin{document}

\title{Towards Interpretable Counterfactual Generation via Multimodal Autoregression}
\titlerunning{Towards ICG via Multimodal Autoregression}

\author{Chenglong Ma\inst{1,2}\textsuperscript{$\dagger$} \and
Yuanfeng Ji\inst{3}\textsuperscript{$\dagger$} \and 
Jin Ye\inst{4} \and 
Lu Zhang\inst{5}  \and 
Ying Chen\inst{4}  \and \\
Tianbin Li\inst{4} \and
Mingjie Li\inst{3} \and
Junjun He\inst{2,4}\textsuperscript{*} \and
Hongming Shan\inst{1}\textsuperscript{*}}
\authorrunning{C. Ma et al.}

\institute{Fudan University \and
Shanghai Innovation Institute \and
Stanford University \and 
Shanghai AI Laboratory \and 
Jinan University
}

\maketitle 

\let\thefootnote\relax\footnotetext{$^\dagger$Equal contribution.\\$^*$Co-corresponding authors.}
\begin{abstract}
Counterfactual medical image generation enables clinicians to explore clinical hypotheses, such as predicting disease progression, facilitating their decision-making. 
While existing methods can generate visually plausible images from disease progression prompts, they produce \emph{silent} predictions that lack interpretation to verify how the generation reflects the hypothesized progression---a critical gap for medical applications that require traceable reasoning. In this paper, we propose \textbf{I}nterpretable \textbf{C}ounterfactual \textbf{G}eneration (\mytask),  
a novel task requiring the joint generation of counterfactual images that reflect the clinical hypothesis and interpretation texts that outline the visual changes induced by the hypothesis. 
To enable \mytask, we present \datasetname, the first dataset pairing longitudinal medical images with hypothetical progression prompts and textual interpretations. 
We further introduce \methodname, an autoregressive model that unifies the generation of counterfactual images and textual interpretations. We demonstrate the superiority of \methodname in generating progression-aligned counterfactuals and interpretations, showing significant potential in enhancing clinical decision support and medical education. Project page: \url{https://progemu.github.io}.


\end{abstract}

\section{Introduction}
\label{sec:intro}
Counterfactual generation, which aims to synthesize hypothetical ``what-if'' scenarios based on real data, has emerged as a transformative tool for medical image analysis~\cite{chambon2022roentgen,lee2024clinical}. 
By generating realistic counterfactual images from a patient's scan under various clinical scenarios (\eg, disease progression and regression), clinicians can visualize potential outcomes and customize patient-specific interventions, thereby enhancing decision-making and medical education.

Recent advances in generative models have shown promising results in counterfactual medical image generation, particularly in chest X-ray (CXR) analysis~\cite{chambon2022roentgen,gu2023biomedjourney,liang2023pie}.
Although these models can generate visually plausible CXR images from input images and disease progression prompts (\eg, ``New small left pleural effusion''),  their clinical application is limited by what we call \emph{silent} predictions.
In other words, the models produce counterfactual images without providing accompanying texts that explain the specific radiological changes associated with the disease progression (\eg, ``The left hemidiaphragm shows new blunting in the posterior pleural sulcus, indicating a new small left pleural effusion''). 
Consequently, these models fail to establish a transparent, causal chain to validate the counterfactual generation, undermining their utility in educational or decision-support applications that require traceable reasoning.

To this end, we introduce \textbf{I}nterpretable \textbf{C}ounterfactual \textbf{G}eneration (\mytask), a novel task that requires the joint generation of a counterfactual image reflecting the input hypothesis and an interpretation text justifying the generated changes, thereby ensuring causal alignment between the produced images and texts.
However, \mytask presents two key challenges, which can be summarized as follows: (a) \textit{data scarcity}--publicly available datasets rarely include longitudinal studies with paired interpretation for the changes observed between prior and subsequent images, and  
(b) \textit{disjointed modeling}--existing solutions typically isolate image and text generation~\cite{gu2023biomedjourney,kyung2024ehrxdiff,li2024contrastive,shentu2024cxrirgen}, hindering effective modeling of vision-language dependencies. 

In this paper, we tackle \mytask through two key innovations. 
First, we focus on CXR modality and curate the \datasetname dataset using a GPT4-based pipeline (Fig.~\ref{fig:dataset}).
To our knowledge, \datasetname is the first large-scale longitudinal CXR dataset designed for interpretable counterfactual generation.
It contains 11,439 quadruples from 7,388 unique patients, each including a prior image, a subsequent image, a progression prompt, and a corresponding textual interpretation of the observed changes.
Second, we develop \methodname, a multimodal autoregressive model that jointly generates counterfactual images and their corresponding textual interpretations (Fig.~\ref{fig:model}). By unifying these modalities, \methodname ensures that the generated image aligns with the hypothesized progression while the text outlines the underlying radiological changes. 
Our contributions can be summarized as follows: 
\begin{itemize}
    \item We formalize \mytask as a novel task requiring the joint generation of counterfactual images and interpretation texts. 
    \item We curate the \datasetname dataset, addressing the critical data scarcity for the \mytask task.
    \item We introduce \methodname, a multimodal autoregressive model that unifies image and text generation for interpretable counterfactual generation.
    \item Quantitative and qualitative results show that \methodname achieves state-of-the-art generation of progression-aligned counterfactual images and interpretation texts, highlighting its potential for counterfactual medical analysis.
\end{itemize}

\section{Related Work}
\label{sec:related}

\noindent \textbf{Counterfactual Medical Image Generation.}
Counterfactual medical image generation shows significant potential for simulating hypothetical disease scenarios~\cite{alaya2024mededit,gu2023biomedjourney,kyung2024ehrxdiff,liang2023pie,shentu2024cxrirgen} and improving the interpretability of AI diagnostic models~\cite{cohen2021gifsplanation,fang2024decoding,fang2024diffexplainer,liu2024controlcg}.
BiomedJourney~\cite{gu2023biomedjourney} frames the counterfactual image generation problem as an image-editing task and adapts an image-editing framework~\cite{brooks2023instructpix2pix} to patient journey data for generating counterfactual images.
PIE~\cite{liang2023pie} uses diffusion-based~\cite{rombach2022sd} progressive image editing to simulate realistic disease progression, although it relies on external region masks and does not explain the generated images.
CXR-IRGen~\cite{shentu2024cxrirgen} combines a diffusion model with a large language model~\cite{lewis2020bart} to produce paired CXR images and radiology reports simultaneously, yet the text only describes the counterfactual image without highlighting differences between images.
In contrast, our  \methodname unifies the generation of counterfactual images and interpretation texts within a single model.

\noindent \textbf{Unified Multimodal Generative Models.}
Recent breakthroughs in generative AI have enabled single model to  generate various data types (\eg, image and text) simultaneously~\cite{chen2025januspro,team2024chameleon,wang2024emu3}, achieving superior performance on multiple tasks.
Chameleon~\cite{team2024chameleon} discretizes visual and textual data into tokens and trains a generic multimodal transformer from scratch in an autoregressive fashion, enabling seamless generation of images and texts. 
Emu3~\cite{wang2024emu3} extends this approach and enhances the generation of images, texts, and videos by incorporating a more powerful vision tokenizer and an improved post-training strategy. 
Following this line of work, we propose an autoregressive model, \methodname, that leverages the strengths of unified vision-language processing and autoregressive modeling, making it naturally suitable for the proposed interpretable counterfactual generation task.

\section{Methodology}
\label{sec:method}
For a consecutive study pair, given a prior image and a disease progression prompt, the \mytask task aims to generate a progression-aligned counterfactual image and an interpretation text describing the progression-induced radiological changes between prior and counterfactual images. 
We address this task by first curating the \datasetname dataset and then developing an autoregressive multimodal generative model, \methodname, detailed below.

\subsection{\datasetname Dataset Curation}

CXR image-report datasets like MIMIC-CXR~\cite{johnson2019mimic} (377,110 pairs from 227,827 studies) and CheXpertPlus~\cite{chambon2024chexpertplus} (223,228 pairs from 187,711 studies) are widely used in medical imaging research.  
However, these datasets are not directly applicable for the interpretable counterfactual generation task, because they lack comparative descriptions between two CXR studies, and a non-trivial number of studies are of low quality, requiring further processing. 

To this end, we curate \datasetname, a longitudinal dataset for the interpretable counterfactual generation task, derived from MIMIC-CXR and CheXpertPlus.
We start the curation by keeping only those with posterior-anterior view images and discarding any with empty ``Impression'' or ``Findings'' sections. 
For each patient, we form study pairs by selecting a prior CXR study and a subsequent one conducted within 100 days to minimize unrelated confounding factors, and then resize the images to a 512×512 resolution and spatially align them using SimpleITK's registration.

\begin{figure}[t]
    \centering
    \includegraphics[width=\linewidth]{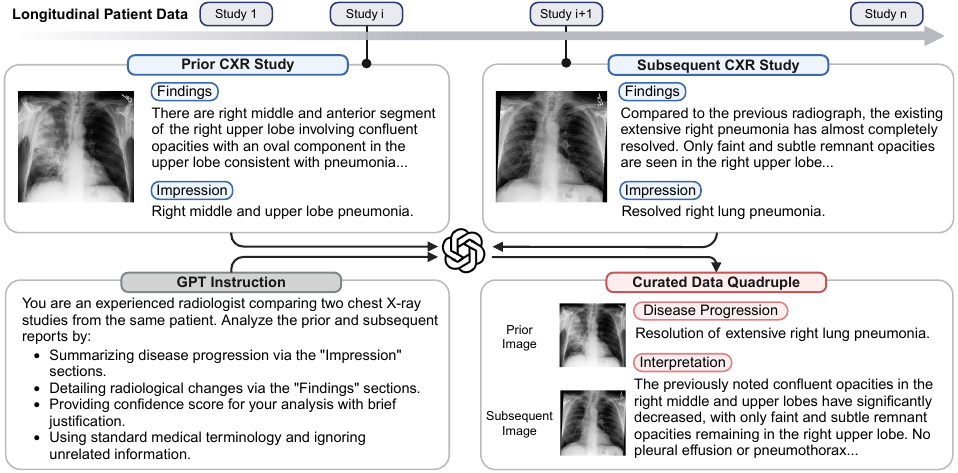}
    \caption{Dataset curation process of \datasetname. Two consecutive CXR studies and their corresponding ``Findings'' and ``Impression'' sections are fed into GPT along with instructions, generating a disease progression prompt and an interpretation text. The resulting data quadruple is curated for the interpretable counterfactual generation.}
    \label{fig:dataset}
\end{figure}

Following this, we employ GPT4 to generate disease progression prompts and interpretation texts. 
As illustrated in Fig.~\ref{fig:dataset}, for each study pair, GPT4 compares the ``Impression'' sections of the prior and subsequent reports to summarize disease progression and contrasts the ``Findings'' sections to detail the radiological differences underlying the visual changes observed between the CXRs, while also providing a confidence score based on the clarity and completeness of the reports.
As a result, each curated data quadruple in \datasetname includes the prior and subsequent CXRs, the disease progression prompt, and the corresponding interpretation text.

After filtering out quadruples with low confidence scores or issues such as poor image quality or spatial misalignment, the final curated dataset comprises 11,439 data quadruples from 7,388 patients, where 7,784 quadruples are sourced from MIMIC-CXR and 3,655 are from CheXpertPlus. The dataset is then split into training and test sets, with 10,679 quadruples used for training and 760 quadruples reserved for testing, ensuring no overlap between patients.

\begin{figure}[tb]
    \centering
    \includegraphics[width=\linewidth]{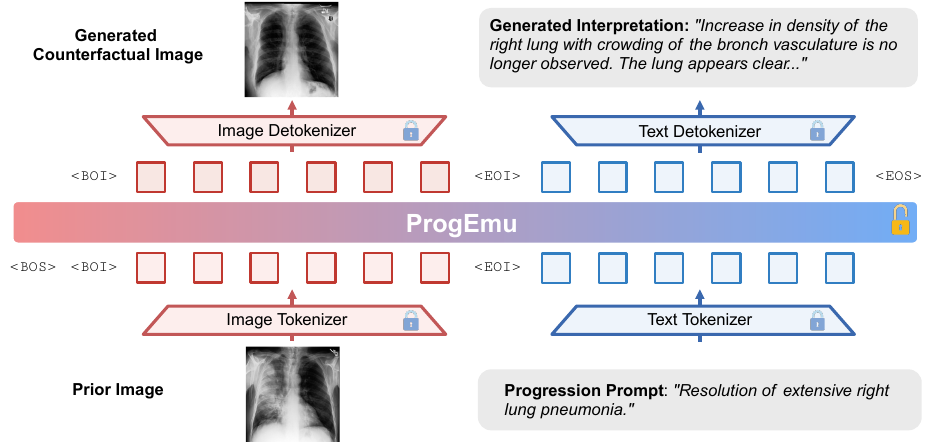}
    \caption{Overview of \methodname , an autoregressive model for interpretable counterfactual generation. The prior image is tokenized along with the textual progression prompt, and the model performs next-token prediction to generate both image and text tokens. These tokens are then detokenized to produce the counterfactual image and corresponding interpretation.}
    \label{fig:model}
\end{figure}

\subsection{Multimodal Autoregression for \mytask}
\mytask demands causal alignment between prior and subsequent data and joint modeling of image and text modalities. 
Existing approaches either do not produce interpretation texts or isolate image generation and text generation~\cite{fang2024decoding,li2024contrastive,shentu2024cxrirgen}, leading to fragmented reasoning and unexplored multi-modality synergy. 
To this end, we introduce a multimodal autoregressive transformer, \methodname.

As illustrated in Fig.~\ref{fig:model}, \methodname integrates both image and text modalities in a unified framework to jointly generate counterfactual images and corresponding interpretation texts. 
The model begins by tokenizing the prior image, the disease progression prompt, the subsequent image, and the interpretation text into respective discrete token sequences, creating a mixed-modal token sequence formatted as ``\texttt{<BOS><BOI>}\{prior image tokens\}\texttt{<EOI>}\{prompt tokens\}\texttt{<BOI>}\{subsequent image tokens\}\texttt{<EOI>}\{interpretation text tokens\}\texttt{<EOS>}'', where \texttt{<BOS>} and \texttt{<EOS>} are predefined special tokens indicating the beginning and end of the entire sequence, and \texttt{<BOI>} and \texttt{<EOI>} separate the modalities. 

During training, this mixed-modal token sequence is fed into \methodname. 
The model observes the input part (\ie, prior image tokens and progression prompt tokens) and is tasked with predicting the target part (\ie, subsequent image tokens and interpretation text tokens) in an autoregressive manner, minimizing the cross-entropy loss between its generated token and the target token at each step in the sequence:
\begin{align}
    \mathcal{L} = -\sum_i \log p_\theta (v_i|u_1,...,u_N,v_1,...,v_{i-1}),
    \label{eq:loss}
\end{align}
where $\theta$ denotes the parameters of \methodname, and $u_*$ and $v_*$ denote tokens in the input and target parts, respectively. By representing visual and textual data as discrete tokens, \methodname seamlessly generates counterfactual images and interpretation texts within a single model, and its autoregressive nature ensures causality between the input-counterfactual image pair and the generated interpretation, suitable for the \mytask task.

\section{Experiments}
\label{sec:exp}

\begin{table}[!tbp]
\centering
\caption{Quantitative evaluation for state-of-the-art methods on the \datasetname dataset. The best results are highlighted in \textbf{bold} and the second best results are \underline{underlined}. \methodnamesep: a cascade variant of \methodname for ablation study. %``$\uparrow$'': the higher the better. ``$\downarrow$'': the lower the better.
}
\resizebox{1.0\textwidth}{!}{% <------ Don't forget this %
\begin{tabular}{lccccccc}
\toprule
{} & \multicolumn{4}{c}{Counterfactual Image} & \multicolumn{3}{c}{Explanatory Text} \\
\cmidrule(lr){2-5} 
\cmidrule(lr){6-8}
{\textbf{Method}} & FID$\downarrow$ & CLIP$\uparrow$ & AUROC$\uparrow$ & F1$\uparrow$ & BLEU-3$\uparrow$ & METEOR$\uparrow$ & ROUGE-L$\uparrow$ \\
\midrule
PIE~\cite{liang2023pie} 
&37.60 &34.03 
&0.7735 &\underline{0.8894} 
&-- &-- &-- \\
BiomedJourney~\cite{gu2023biomedjourney} 
&36.62 &34.59 
&\textbf{0.8385} &0.8411 
&-- &-- &-- \\
CXR-IRGen~\cite{shentu2024cxrirgen} 
&35.39 &32.51 
&0.5236 &0.7609 
&0.0448 &0.2115 &0.1846 \\
\midrule
\methodnamesep (\textbf{ours}) 
&\underline{29.51} &\underline{35.00} 
&\underline{0.7951} &0.8853 
&\textbf{0.2492} &\textbf{0.5660} &\underline{0.2496} \\
\methodname (\textbf{ours}) 
&\textbf{29.21} &\textbf{35.24} 
&0.7921 &\textbf{0.8914} 
&\underline{0.1241} &\underline{0.4097} &\textbf{0.2606} \\
\bottomrule
\end{tabular}
}
\label{tab:comparison}
\end{table}

\subsection{Experimental Setup}
\textbf{Implementation Details.} Training a well-performing multimodal model from scratch demands significant computational costs~\cite{shi2024llamafusion}. Instead, we reuse the autoregressive multimodal model, Emu3~\cite{wang2024emu3}, to initialize the weights of \methodname, and adopt its image tokenizer that encodes a $512 \times 512$ image into $4,096$ discrete tokens. We train \methodname for 30 epochs with a batch size of 8 on 4 NVIDIA H100 GPUs using AdamW~\cite{loshchilov2018adamw} optimizer. The base learning rate is $10^{-5}$, following a cosine annealing with minimum learning rate of $10^{-6}$.

\noindent \textbf{Evaluation Metrics.} For the generated counterfactual images, we evaluate their visual quality, prompt alignment, and preservation of pathological information. 
Visual quality is assessed using Fr\'echet Inception Distance (FID) with Inception-v3~\cite{heusel2017fid} and a pathology classifier~\cite{cohen2020xrv} pretrained on multiple CXR datasets. The alignment with the prompt is measured by CLIP score using BiomedCLIP~\cite{zhang2023biomedclip}. For pathological information preservation, following BiomedJourney~\cite{gu2023biomedjourney}, we apply the pathology classifier to the counterfactual images, comparing predicted pathologies with labels from radiology reports in the subsequent CXR study, and calculate the area under the Receiver Operating Curve (AUROC) and F1 score. 
For the generated interpretation texts, we use BLUE-3~\cite{papineni2002bleu}, METEOR~\cite{banerjee2005meteor} and ROUGE-L~\cite{lin2004rouge} scores, which are widely applied in text generation evaluation.

\subsection{Comparison with State-of-the-Art}
We compare \methodname to three state-of-the-art methods for counterfactual generation, including PIE~\cite{liang2023pie}, BiomedJourney~\cite{gu2023biomedjourney}, and CXR-IRGen~\cite{shentu2024cxrirgen}. 
To adapt these models to the new \mytask task, we finetune them on the training data of \datasetname dataset. The comparison results are presented in Table~\ref{tab:comparison}.

While BiomedJourney and PIE show competitive image generation capability, they do not provide textual interpretation for the counterfactual image, limiting their utility in clinical applications. CXR-IRGen achieves competitive FID, but the AUROC and F1 score are relatively low. This may be attributed to its architecture, where the large language model tries to generate interpretation texts from solely the counterfactual image without considering the prior one, forcing the counterfactual embedding to retain features from the prior image and ultimately leading to a loss of pathological features in the counterfactual image. 
In contrast, by unifying image and text generation into one single model, our proposed \methodname demonstrates superiority over state-of-the-art counterfactual generation methods in \mytask task, providing understandable insights for the predicted disease progression. 

\begin{figure}[t]
    \centering
    \includegraphics[width=\linewidth]{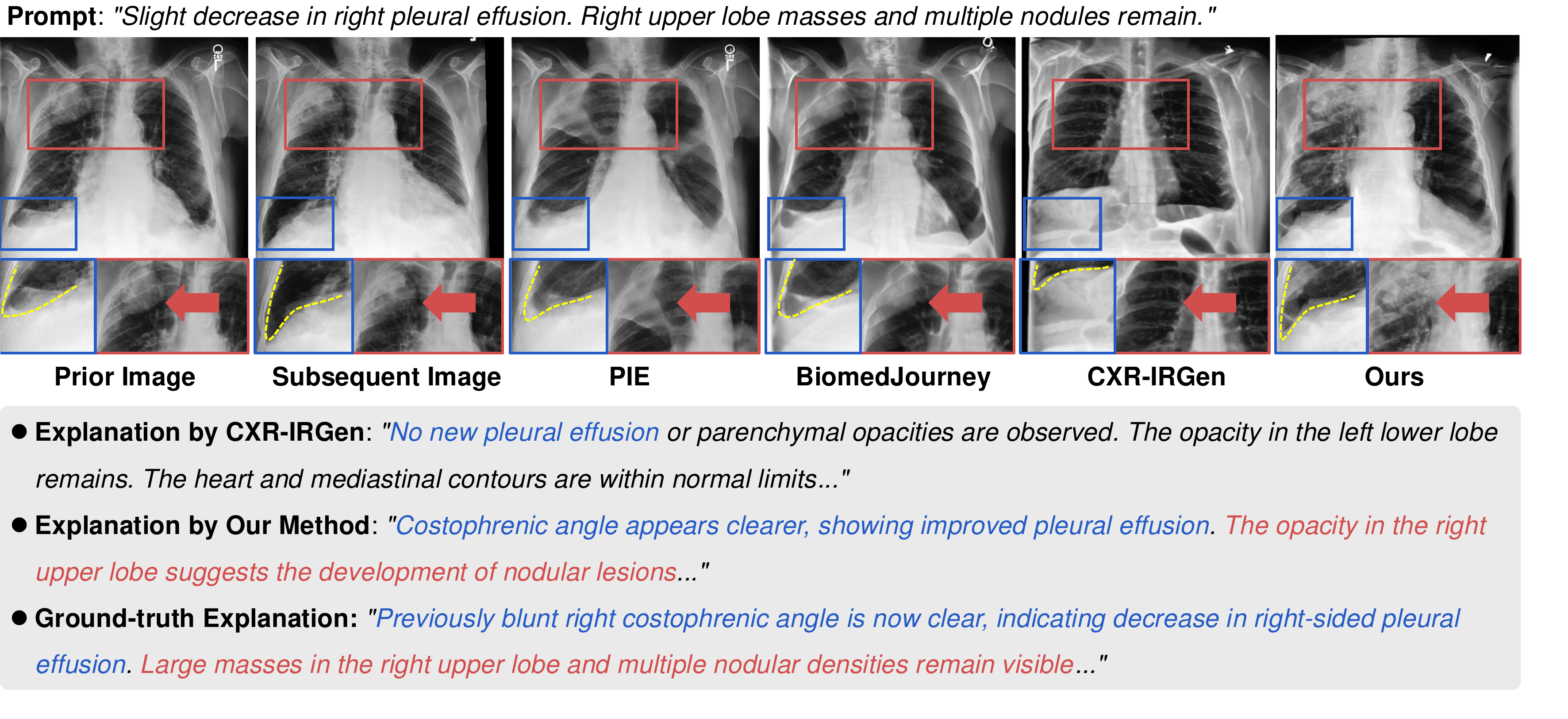}
    \caption{Qualitative comparison of counterfactual images generated by different methods with enlarged regions of interest. For each image, the right costophrenic angle is outlined by a yellow dashed curve, and the masses and nodules in the right upper lobe are indicated by a red arrow.}
    \label{fig:sample1}
\end{figure}

Fig.~\ref{fig:sample1} provides a qualitative example, where the prior image shows pleural effusion and nodules in the right upper lobe. Given the prompt ``Slight decrease in right pleural effusion. Right upper lobe masses and multiple nodules remain'', \methodname generates a counterfactual image that closely matches the real subsequent image, where the decrease of pleural effusion is evidenced by a sharper right costophrenic angle. The nodules indicated by the opacities in the right upper lobe, which are blurred or removed in the images generated by other methods, remain clearly visible in \methodname's output. Notably, the radiological changes suggestive of the disease progression are also well captured in the interpretation texts generated by \methodname. 

\begin{figure}[t]
    \centering
    \includegraphics[width=\linewidth]{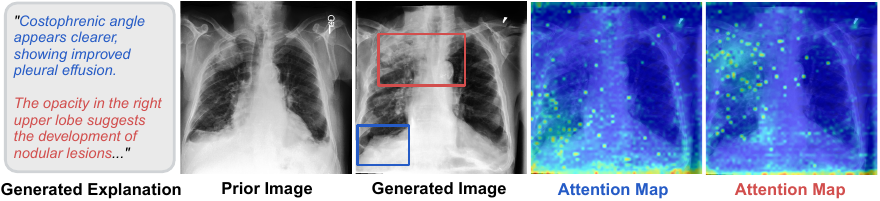}
    \caption{Cross-attention maps between image and text tokens generated by \methodname. The generated interpretation (left) includes color-coded phrases (``Costophrenic angle...'' in \textcolor[HTML]{255BC6}{blue}, ``The opacity...'' in \textcolor[HTML]{D24E4D}{red}). The bounding boxes on the generated image (center) match these phrases, and the attention maps (right) show how the model aligns text tokens with the corresponding region.}
    \label{fig:attn}
\end{figure}

To further investigate whether \methodname internally learns to explain the counterfactuals, we visualize the cross-attention maps between generated image tokens and text tokens in Fig.~\ref{fig:attn}, where the region highlighted in each attention map is consistent with the generated texts. This demonstrates \methodname's ability to explain the radiological changes in the generated image, establishing a causal chain for interpretable counterfactual generation.

\subsection{Ablation Study}
To better understand the contribution of unified image and text generation, we develop \methodnamesep, a cascade of two networks: an ``image-only'' network that generates only the counterfactual image without interpretation, and a ``text-only'' network that generates only the interpretation text. Both networks share the same architecture as \methodname and are trained independently with the same dataset and setting. 

During inference, the image-only network predicts the counterfactual image, which will be fed into the text-only network along with the prior image and progression prompt for generating the textual interpretation. 
As shown in the last two rows of Table~\ref{tab:comparison}, \methodnamesep achieves only competitive performance compared to \methodname, despite doubling the model size. This suggests that the joint modeling of image and text (as in \methodname) may offer synergistic benefits and facilitate the \mytask task.

\section{Conclusion}
\label{sec:concl}

We introduce interpretable counterfactual generation (\mytask), a novel task aimed at producing progression-aligned counterfactual images with textual interpretation for the progression-induced radiological changes. To enable \mytask, we curate the \datasetname dataset, the first longitudinal CXR dataset annotated with progression prompts and explanations. Using this dataset, we develop \methodname, a multimodal autoregressive model that generates interpretable counterfactuals. Experimental results demonstrate that \methodname outperforms existing counterfactual generation methods, 
highlighting its potential for advancing counterfactual medical image analysis. The \datasetname dataset will be made publicly available. Other future directions include generalizing to more imaging modalities and conducting real-world clinical validation.

\bibliographystyle{splncs04}
\bibliography{ref-cg}




% LNCS Homepage, \url{http://www.springer.com/lncs}, last accessed 2023/10/25
% \end{thebibliography}
\end{document}